\documentclass{PoS}

\title{Three-loop vertex corrections}

\ShortTitle{Three-loop vertex corrections}

\author{\speaker{
  Matthias Steinhauser
  }%
         \thanks{SFB/CPP-10-07, TTP10-02}\\
  \\
        Karlsruhe Institute of Technology\\
        E-mail: \email{matthias.steinhauser@kit.edu}}

\abstract{We describe recent evaluations of three-loop vertex
  corrections which have been performed in the context of different
  physical applications: the massless quark and gluon form factor, the
  vector current matching coefficient between QCD and NRQCD,
  and the virtual corrections of the gluon-Higgs coupling with
  finite top quark mass.}

\FullConference{RADCOR 2009 - 9th International Symposium on Radiative
Corrections (Applications of Quantum Field Theory to Phenomenology) \\
                 October 25-30 2009\\
                 Ascona, Switzerland}

\begin{document}


\newcommand{\conepm}{X_{9,1}}
\newcommand{\ctwop}{X_{9,2}}
\newcommand{\cfourp}{X_{9,4}}

\section{\label{sec::3lff}Three-loop fermion and gluon form factors}

\begin{figure}[b]
  \begin{center}
    \begin{tabular}{ccc}
      \includegraphics[width=7em]{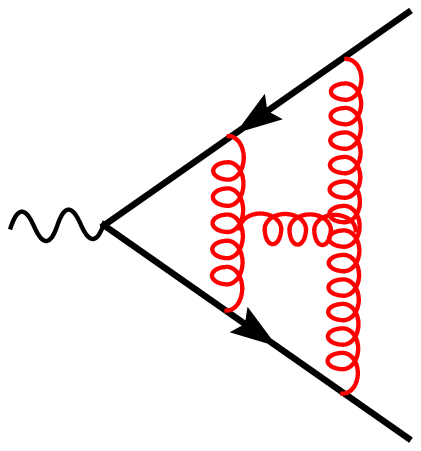} 
      &
      \mbox{}\hspace*{3em}
      \includegraphics[width=7em]{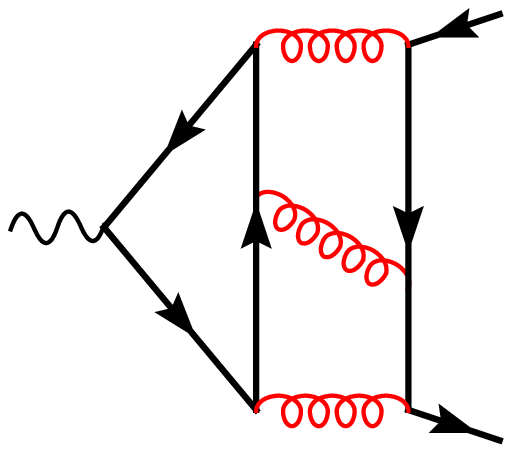} 
      \hspace*{3em}\mbox{}
      &
      \includegraphics[width=7em]{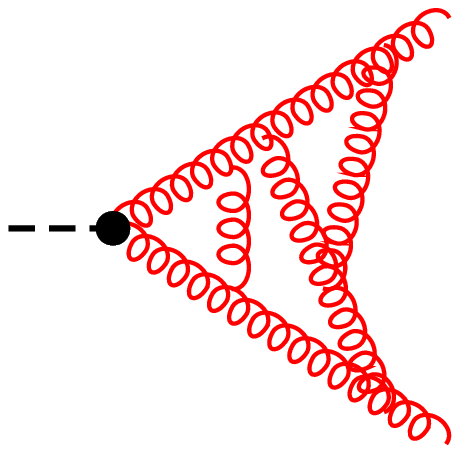}
      \\
      (a) & (b) & (c)
    \end{tabular}
    \caption{\label{fig::diags}Sample Feynman diagrams contributing to the
      $F_q$ ((a) and (b)) and $F_g$ (c)
      at three-loop order. Straight and curly lines denote
      quarks and gluons, respectively.}
  \end{center}
\end{figure}

In this Section we consider massless QCD and discuss the
virtual corrections to the photon-quark and Higgs-boson-gluon vertex
(see also Ref.~\cite{Todtli:2008bn} and references therein). 
It is convenient to decompose the vertex functions according to the
Lorentz structure and define the form factors $F_q$ and $F_g$ via
$  \Gamma^\mu_q = \gamma^\mu F_q(q^2)
  \,,
  \Gamma^{\mu\nu}_g \,\,=\,\, \left(q_1\cdot q_2\,\,
    g^{\mu\nu}-q_{1}^{\nu}\,q_{2}^{\mu}\right)
  F_g(q^2)
  \,,
$
where
$q=q_1+q_2$ and $q_1$ ($q_2$) is the incoming (anti-)quark momentum in
the case of $F_q$, and $F_g$ depends on the gluon momenta $q_1$ and $q_2$ with
polarization vectors $\varepsilon^{\mu}(q_1)$ and $\varepsilon^{\nu}(q_2)$.
Both for $F_q$ and $F_g$, which are obtained by applying appropriate
projectors, we have $q_1^2=q_2^2=0$.
Some sample Feynman diagrams contributing to $F_q$ and $F_g$ are shown in
Fig.~\ref{fig::diags}.

For our calculation we have used two different setups.
The basic idea of the first one has been described in
Refs.~\cite{Baikov:1996rk}: integral
representations for the coefficients of the master integrals 
are derived. They depend on the exponents
of the denominators of the integral under consideration and the
space-time dimension $d$. In the recent years
a procedure has been developed to evaluate the resulting parameter
integrals in the limit of large $d$ (see, e.g.,
Ref.~\cite{Baikov:2007zza}). Knowing sufficiently many expansion
terms the coefficient function can be 
reconstructed since (for fixed exponents) it is a rational function in $d$.
The evaluation of the three-loop vertex corrections profited quite a lot from
the experience gained in the context of the evaluation of the four-loop
two-point functions~\cite{Baikov:2008jh} and the findings of
Ref.~\cite{Baikov:2000jg}. In the 
latter paper it has been shown that the recurrence relations of $n$-loop
three-point functions are equivalent to $(n+1)$-loop two-point functions.

The second method has only been applied to the singlet diagrams (see,
e.g., Fig.~\ref{fig::diags}(b)) contributing to
$F_q$. It relies on the idea to combine the Laporta
method~\cite{Laporta:1996mq} with the Gr\"obner 
bases technique~\cite{S2} which has been published in the computer code {\tt
  FIRE}~\cite{FIRE}.

We parameterize the results for $F_q$ and $F_g$ in terms of the bare coupling
which allows us to 
factorize all occurring logarithms of the form $\ln(Q^2/\mu^2)$, where
$Q^2=-q^2>0$, and to cast the expressions in the form ($x=q,g$)
\begin{eqnarray}
  F_x &=& 1 + \sum_n \left(\frac{\alpha_s}{4\pi}\right)^n
  \left(\frac{\mu^2}{Q^2}\right)^{n\epsilon} F_x^{(n)}
  \,.
\end{eqnarray}
We refrain from listing the results in terms of general
SU(3) colour factors, which can be found in Ref.~\cite{Baikov:2009bg},
however, we present for illustration the finite\footnote{We refer to
  Refs.~\cite{Moch:2005id} for the divergent contribution.} 
part of $F_g$ in the case of QCD
where it takes the form
\begin{eqnarray}
  F_g^{(3)} &=&
  \frac{14423912}{243}
  + \frac{384479\zeta_2}{108}
  - \frac{370649\zeta_3}{18}
  + \frac{280069\zeta_4}{32}
  + \frac{49167\zeta_2\zeta_3}{4}
  - \frac{199263\zeta_5}{10}
  + \frac{1635\zeta_3^2}{4}
  \nonumber\\&&\mbox{}
  - \frac{4527765\zeta_6}{256}
  - 27\conepm + 54\ctwop 
  + n_f\Bigg(-\frac{11801309}{1458} - \frac{42296\zeta_2}{81} + \frac{41018\zeta_3}{81} - 
  247\zeta_2\zeta_3 
  \nonumber\\&&\mbox{}
  + \frac{1055\zeta_4}{8} + \frac{16982\zeta_5}{45}
  \Bigg) 
  + n_f^2\Bigg(\frac{2239573}{4374} + \frac{4\zeta_2}{3} + \frac{3376\zeta_3}{81} + \frac{349\zeta_4}{18}
  \Bigg) 
  \,.
  \label{eq::FqFg}
\end{eqnarray}
The two constants $X_{9,1}$ and $X_{9,2}$ 
take the numerical values
$X_{9,1}\approx1428.9963678666183591$ and
$X_{9,2}\approx528.0583  \pm  0.0326$
where $X_{9,1}$ is available analytically~\cite{Heinrich:2009be} and
$X_{9,2}$ is known
numerically~\cite{Baikov:2009bg,Heinrich:2009be} with the indicated
precision. 

The new NNNLO results for the form factors
constitute building blocks for a number of applications. Among them are the
virtual corrections to Higgs boson production in gluon fusion, the
Drell-Yan process and the two-jet cross section in $e^+ e^-$ collisions.


\section{\label{sec::cv}Three-loop matching of the vector current}

In contrast to the previous Section we consider here QCD with $n_h=1$ massive
and $n_l$ massless quarks and evaluate the matching coefficient between QCD
and non-relativistic QCD (NRQCD) for the vector current. This quantity is
important for phenomena where two heavy quarks are produced in
electron-positron annihilation or a bound state of two heavy quarks
decays into a lepton pair.

The vector currents in QCD and NRQCD are given by
  $j_v^\mu = \bar{Q} \gamma^\mu Q,$ and
  $\tilde{j}^k = \phi^\dagger \sigma^k \chi,$  
where $Q$ denotes a generic heavy quark with mass $m_Q$ and $\phi$ and
$\chi$ are two-component Pauli spinors for quark and
anti-quark, respectively, and $\sigma^k$ ($k=1,2,3$) are the Pauli
matrices. The two currents $j_v^\mu$ and $\tilde{j}^k$ can be used to
compute vertex corrections with two on-shell quarks and momenta $q_1$
and $q_2$ ($\Gamma_v$ and $\tilde{\Gamma}_v$). From the
requirement that the results agree up to power-corrections in $m_Q$
defines the matching coefficient $c_v$
\begin{eqnarray}
  Z_2 \Gamma_v &=& c_v \tilde{Z}_2 \tilde{Z}_v^{-1} \tilde{\Gamma}_v
  + \ldots
  \,,
  \label{eq::def_cv}
\end{eqnarray}
where $Z_2$ denotes the on-shell wave function renormalization
constant~\cite{Broadhurst:1991fy,Melnikov:2000zc}
and quantities with a tilde are defined within NRQCD.
The ellipses in Eq.~(\ref{eq::def_cv}) represent terms of order $1/m_Q$
which are neglected.

For the evaluation of $c_v$ it is convenient to consider
$q^2=(q_1+q_2)^2\approx 4 m_Q^2$ and apply the so-called threshold
expansion~\cite{Beneke:1997zp} to $\Gamma_v$ which
identifies the hard, 
soft, potential and ultra-soft integration regions. 
The latter three contributions are present both on the left- and 
right-hand side of Eq.~(\ref{eq::def_cv}) and thus cancel out. Only the hard
contribution where $q^2=4m_Q^2$ and which is only present in $\Gamma_v$ 
has to be evaluated.
This reasoning is based on the use of Dimensional Regularization which
is crucial for the evaluation of higher order corrections to $c_v$
since scaleless integrals are automatically set to zero. This concerns
in particular $\tilde{Z}_2$ and $\tilde{\Gamma}_v$ which are both
identical to one.

The two-loop corrections to $c_v$ have been evaluated in
Refs.~\cite{Czarnecki:1997vz,Beneke:1997jm,Kniehl:2006qw} 
and the fermionic three-loop
contribution in Refs.~\cite{Marquard:2006qi,Marquard:2009bj}.
The setup used in Ref.~\cite{Marquard:2009bj}
is completely automated. It is based on a chain of
programs which work hand-in-hand. The starting point is {\tt
  QGRAF}~\cite{Nogueira:1991ex} which generates the
amplitudes for each 
Feynman diagram. In a next step {\tt q2e} and {\tt
  exp}~\cite{Harlander:1997zb} are used in order
to identify the topologies and generate input expression in {\tt
  FORM}~\cite{Vermaseren:2000nd} format. The same input file
containing the 
description of the topologies needed by {\tt exp} is
also used in order to provide the input for {\tt crusher}~\cite{PMDS}
which performs the reduction of all occurring integrals to a basic
set, so-called the master integrals. The topology file is also
used for providing the necessary input
for {\tt FIESTA}~\cite{Smirnov:2008py,Smirnov:2009pb} which 
is employed for the numerical evaluation of all master integrals.

There are several checks on the correctness of our results. 
As far as the analytical part is concerned 
the most important one is the evaluation of the Feynman diagrams for
general gauge parameter $\xi$: we have checked that the linear $\xi$-term
drops out in our three-loop result.
A strong check of the numerical part of our calculation is the change
of the master integral basis, which is achieved analytically with the
help of the integration-by-parts relations generated by {\tt crusher}. In the
new expression, which is a completely different linear combination of master
integrals, the numerical evaluation is again performed with the
help of {\tt FIESTA}.

In numerical form the result for $c_v$ is given by (for $\mu=m_Q$):
\begin{eqnarray}
  c_v &=& 1 - 2.67 \frac{\alpha_s}{\pi}
  + \left(-44.55 + 0.41 n_l\right) \left(\frac{\alpha_s}{\pi}\right)^2
  \nonumber\\
  &&\mbox{}+ \left(c_{v,g} - 0.93(8) n_h - 0.09 n_h n_l 
  + 120.75(1) n_l - 0.82 n_l^2\right) 
  \left(\frac{\alpha_s}{\pi}\right)^3
  + {\cal O}(\alpha_s^4)
  \,.
\end{eqnarray}
Once the purely gluonic contribution $c_{v,g}$ is available, which is expected
to be numerically dominant, $c_v$ can be used in the analysis of the
third-order cross section $\sigma(e^+e^-\to t\bar{t}+X)$ close to
threshold or the extraction of the bottom quark mass from $\Upsilon$
sum rules.


\section{Virtual NNLO corrections to Higgs production in gluon fusion}

In this Section we go beyond the effective theory which has been used in
Section~\ref{sec::3lff} in order to define $F_g$ and consider the gluon-Higgs
vertex with finite top quark mass. This constitutes an important contribution
to the gluon fusion process which has the largest
production cross section for Higgs bosons both at Tevatron and LHC.
Whereas the NLO corrections are exactly
known~\cite{Dawson:1990zj,Djouadi:1991tka,Spira:1995rr} at NNLO until 
recently only approximations for infinitely heavy top quark have been
available. Recently this gap has been closed and in the
works~\cite{Harlander:2009mq,Pak:2009dg,Harlander:2009my} the NNLO
corrections to the cross section $\sigma(pp\to H+X)$ incorporating the
top quark mass dependence have been evaluated. 
The virtual corrections have been evaluated before in
Refs.~\cite{Pak:2009bx,Harlander:2009bw}. In the following we 
describe in more detail the
computation of Ref.~\cite{Pak:2009bx}.

\begin{figure}
  \centering
  \begin{tabular}{cc}
    \begin{minipage}{13em}
      \includegraphics[width=\linewidth]{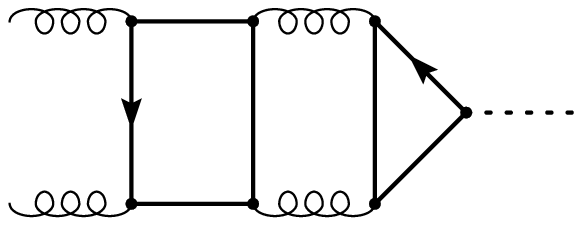}
      \\
      \includegraphics[width=\linewidth]{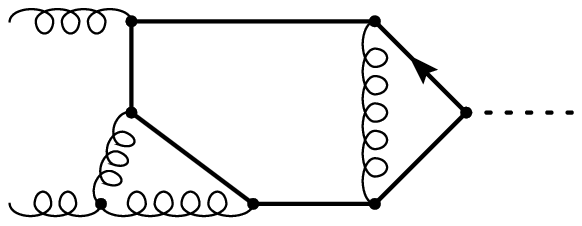}
    \end{minipage}
    &
    \raisebox{-6.5em}{\includegraphics[width=.48\linewidth]{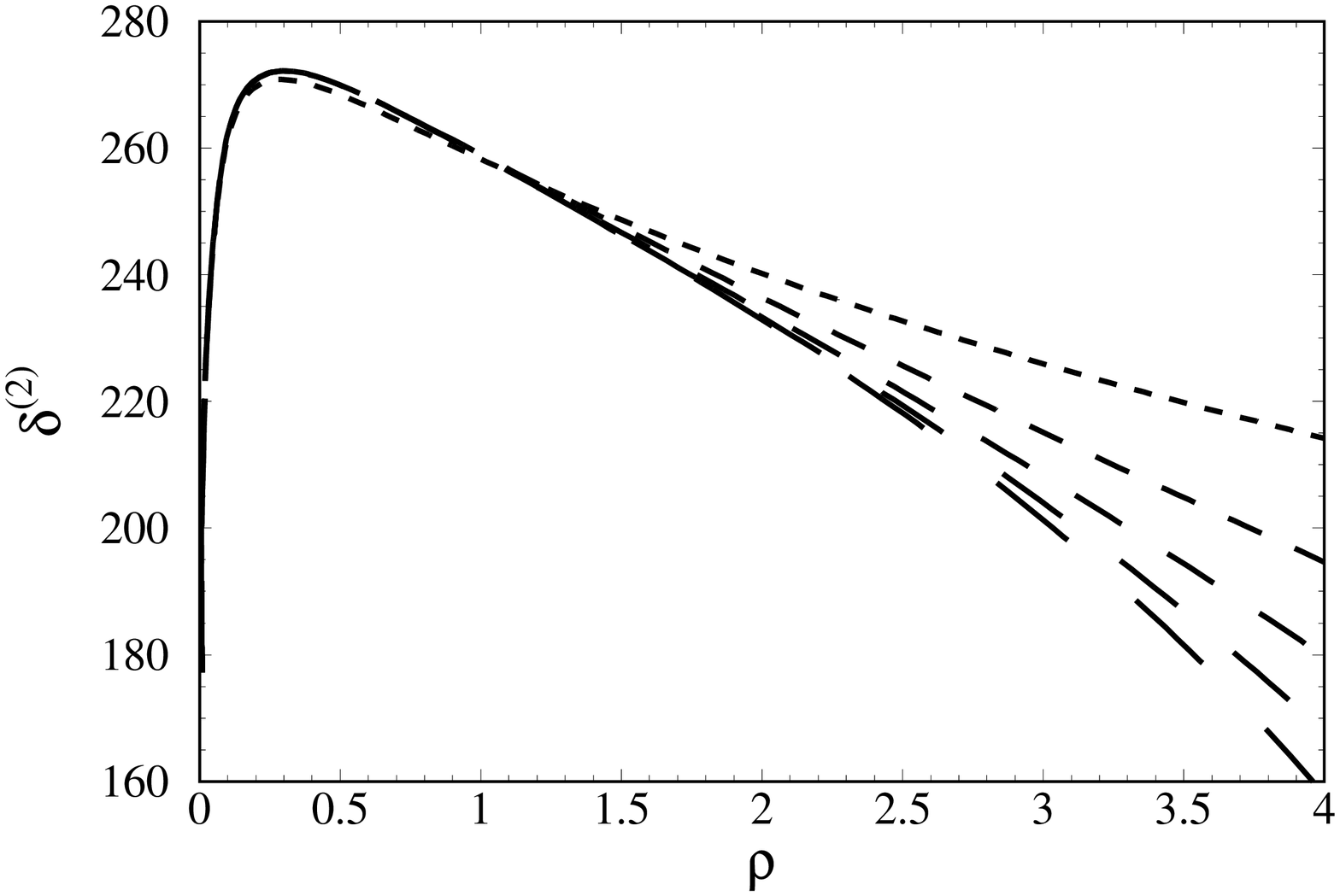}}
    \\
    (a) & (b)
  \end{tabular}
  \caption[]{\label{fig::diag} (a) Sample diagrams contributing to the NNLO
    virtual corrections to $gg\to H$. Solid lines represent the top quark with
    mass $M_t$ and curly (dotted) lines massless gluons (the Higgs boson).
    (b) Finite part of $\delta^{(2)}$
    as a function of $\rho$. The longer-dashed lines include
    successively higher orders in $\rho$.}
\end{figure}

The virtual contribution to the partonic 
cross section can be cast in the form
\begin{eqnarray}
  \hat{\sigma}_{ggh}^{\rm virt} &=& \hat{\sigma}_{\rm LO} \left(
    1 + \frac{\alpha_s}{\pi}~ \delta^{(1)}  
    + \left(\frac{\alpha_s}{\pi}\right)^2 \delta^{(2)} + \ldots
  \right)
  \,,
\end{eqnarray}
where the LO cross section is given by
$  \hat{\sigma}_{\rm LO} =
  {G_F~\alpha_s^2}{f_0(\rho,\epsilon)}~\delta(1-x) /
  ({288\sqrt{2}\pi}{(1-\epsilon)})
  \,,
$
with $x=M_H^2/\hat{s}$.
$\sqrt{\hat{s}}$ is the partonic center-of-mass
energy. The function $f_0$ and the analytical results of the first 
five terms in the $\rho=M_H^2/M_t^2\to0$ expansion for $\delta^{(1)}$ and
$\delta^{(2)}$ can be found in Ref.~\cite{Pak:2009bx}.
Sample diagrams contributing to $\delta^{(2)}$
are shown in Fig.~\ref{fig::diag}(a).

The setup is similar to the one used in Section~\ref{sec::cv}. A difference is
connected to the asymptotic expansion in the limit $M_H^2\ll M_t^2$ which is
performed with the program 
{\tt exp}~\cite{Harlander:1997zb} and independently with
an in-house {\tt Perl} program~\cite{APak_ae}. In this way 
the three-loop vertex integrals are reduced to \mbox{one-,} two-, and three-loop
vacuum integrals which are treated with {\tt
  MATAD}~\cite{Steinhauser:2000ry}, and to massless one- and two-loop
vertex contributions which are reduced to master integrals (see,
e.g., Ref.~\cite{Gehrmann:2005pd}) with
the help of {\tt FIRE}~\cite{Smirnov:2008iw} and an independent 
program based of the Laprota-method.

Explicit results for $\delta^{(1)}$ and $\delta^{(2)}$ can be found in
Ref.~\cite{Pak:2009bx}. In this contribution we discuss the
convergence properties in Fig.~\ref{fig::diag}(b) where the finite part of 
$\delta^{(2)}$ is shown as a function of $\rho$.
One observes good convergence up to $\rho\approx 3$ which corresponds to
$M_H\approx 1.7 M_t\approx 300$~GeV.

\noindent
{\bf Acknowledgments}\\
I would like to thank my colleagues of
Refs.~\cite{Baikov:2009bg,Marquard:2009bj,Pak:2009dg} for the very fruitful collaboration.
This work is supported by DFG through SFB/TR~9.


\begin{thebibliography}{99}

\bibitem{Todtli:2008bn}
  B.~Todtli,
  arXiv:0903.0540 [hep-ph].

\bibitem{Baikov:1996rk}
  P.~A.~Baikov,
  Phys.\ Lett.\ B {\bf 385} (1996) 404,
  arXiv:hep-ph/9603267;
%
   Phys.\ Lett.\  B {\bf 634}, 325 (2006)
   [arXiv:hep-ph/0507053];
%
  V.~A.~Smirnov and M.~Steinhauser,
  Nucl.\ Phys.\  B {\bf 672} (2003) 199
  [arXiv:hep-ph/0307088].

\bibitem{Baikov:2007zza}
  P.~A.~Baikov,
  PoS {\bf RADCOR2007} (2007) 022.

\bibitem{Baikov:2008jh}
  P.~A.~Baikov, K.~G.~Chetyrkin and J.~H.~K\"uhn,
  Phys.\ Rev.\ Lett.\  {\bf 101} (2008) 012002
  arXiv:0801.1821 [hep-ph].

\bibitem{Baikov:2000jg}
  P.~A.~Baikov and V.~A.~Smirnov,
  Phys.\ Lett.\  B {\bf 477} (2000) 367
  [arXiv:hep-ph/0001192].

\bibitem{Laporta:1996mq}
  S.~Laporta and E.~Remiddi,
  Phys.\ Lett.\ B {\bf 379} (1996) 283,
  arXiv:hep-ph/9602417.

\bibitem{S2}
 A.~V.~Smirnov and V.~A.~Smirnov,
 JHEP {\bf 0601}, 001 (2006)
  [arXiv:hep-lat/0509187].

\bibitem{FIRE}
 A.~V.~Smirnov,
  JHEP {\bf 0810}, 107 (2008)
  [arXiv:0807.3243 [hep-ph]].

\bibitem{Baikov:2009bg}
  P.~A.~Baikov, K.~G.~Chetyrkin, A.~V.~Smirnov, V.~A.~Smirnov and
M.~Steinhauser,
  Phys.\ Rev.\ Lett.\  {\bf 102} (2009) 212002
  [arXiv:0902.3519 [hep-ph]].

\bibitem{Moch:2005id}
  S.~Moch, J.~A.~M.~Vermaseren and A.~Vogt,
  JHEP {\bf 0508} (2005) 049
  [arXiv:hep-ph/0507039];
%
  S.~Moch, J.~A.~M.~Vermaseren and A.~Vogt,
  Phys.\ Lett.\  B {\bf 625} (2005) 245
  [arXiv:hep-ph/0508055].

\bibitem{Heinrich:2009be}
  G.~Heinrich, T.~Huber, D.~A.~Kosower and V.~A.~Smirnov,
  Phys.\ Lett.\  B {\bf 678} (2009) 359
  [arXiv:0902.3512 [hep-ph]].

\bibitem{Broadhurst:1991fy}
  D.~J.~Broadhurst, N.~Gray and K.~Schilcher,
  Z.\ Phys.\ C {\bf 52} (1991) 111.

\bibitem{Melnikov:2000zc}
  K.~Melnikov and T.~van Ritbergen,
  Nucl.\ Phys.\ B {\bf 591} (2000) 515,
  arXiv:hep-ph/0005131;
%
  P.~Marquard, L.~Mihaila, J.~H.~Piclum and M.~Steinhauser,
  Nucl.\ Phys.\  B {\bf 773} (2007) 1,
  arXiv:hep-ph/0702185.

\bibitem{Beneke:1997zp}
  M.~Beneke and V.~A.~Smirnov,
  Nucl.\ Phys.\ B {\bf 522} (1998) 321,
  arXiv:hep-ph/9711391;
%
  V.~A.~Smirnov,
  ``Applied asymptotic expansions in momenta and masses,''
  Springer Tracts Mod.\ Phys.\  {\bf 177} (2002) 1.

\bibitem{Czarnecki:1997vz}
  A.~Czarnecki and K.~Melnikov,
  Phys.\ Rev.\ Lett.\  {\bf 80} (1998) 2531,
  arXiv:hep-ph/9712222.

\bibitem{Beneke:1997jm}
  M.~Beneke, A.~Signer and V.~A.~Smirnov,
  Phys.\ Rev.\ Lett.\  {\bf 80} (1998) 2535,
  arXiv:hep-ph/9712302.

\bibitem{Kniehl:2006qw}
  B.~A.~Kniehl, A.~Onishchenko, J.~H.~Piclum and M.~Steinhauser,
  Phys.\ Lett.\ B {\bf 638} (2006) 209,
  arXiv:hep-ph/0604072.

\bibitem{Marquard:2006qi}
  P.~Marquard, J.~H.~Piclum, D.~Seidel and M.~Steinhauser,
  Nucl.\ Phys.\  B {\bf 758} (2006) 144,
  arXiv:hep-ph/0607168.

\bibitem{Marquard:2009bj}
  P.~Marquard, J.~H.~Piclum, D.~Seidel and M.~Steinhauser,
  Phys.\ Lett.\  B {\bf 678} (2009) 269
  [arXiv:0904.0920 [hep-ph]].

\bibitem{Nogueira:1991ex}
  P.~Nogueira,
  J.\ Comput.\ Phys.\  {\bf 105} (1993) 279.

\bibitem{Harlander:1997zb}
  R.~Harlander, T.~Seidensticker and M.~Steinhauser,
  Phys.\ Lett.\ B {\bf 426} (1998) 125,
  arXiv:hep-ph/9712228;
%
  T.~Seidensticker,
  arXiv:hep-ph/9905298.

\bibitem{Vermaseren:2000nd}
  J.~A.~M.~Vermaseren,
  arXiv:math-ph/0010025.

\bibitem{PMDS}
  P.~Marquard and D.~Seidel,
  unpublished.

\bibitem{Smirnov:2008py}
  A.~V.~Smirnov and M.~N.~Tentyukov,
  arXiv:0807.4129 [hep-ph].

\bibitem{Smirnov:2009pb}
  A.~V.~Smirnov, V.~A.~Smirnov and M.~Tentyukov,
  arXiv:0912.0158 [hep-ph].

\bibitem{Dawson:1990zj}
  S.~Dawson,
  Nucl.\ Phys.\  B {\bf 359} (1991) 283.

\bibitem{Djouadi:1991tka}
  A.~Djouadi, M.~Spira and P.~M.~Zerwas,
  Phys.\ Lett.\  B {\bf 264} (1991) 440.

\bibitem{Spira:1995rr}
  M.~Spira, A.~Djouadi, D.~Graudenz and P.~M.~Zerwas,
  Nucl.\ Phys.\  B {\bf 453} (1995) 17,
  arXiv:hep-ph/9504378.

\bibitem{Harlander:2009mq}
  R.~V.~Harlander and K.~J.~Ozeren,
  JHEP {\bf 11} (2009) 088,
  arXiv:0909.3420 [hep-ph].

\bibitem{Pak:2009dg}
  A.~Pak, M.~Rogal and M.~Steinhauser,
  arXiv:0911.4662 [hep-ph].

\bibitem{Harlander:2009my}
  R.~V.~Harlander, H.~Mantler, S.~Marzani and K.~J.~Ozeren,
  arXiv:0912.2104 [hep-ph].

\bibitem{Pak:2009bx}
  A.~Pak, M.~Rogal and M.~Steinhauser,
  Phys.\ Lett.\  B {\bf 679} (2009) 473
  [arXiv:0907.2998 [hep-ph]].

\bibitem{Harlander:2009bw}
  R.~V.~Harlander and K.~J.~Ozeren,
  Phys.\ Lett.\  B {\bf 679} (2009) 467
  [arXiv:0907.2997 [hep-ph]].

\bibitem{APak_ae}
  A.~Pak, unpublished.

\bibitem{Steinhauser:2000ry}
  M.~Steinhauser,
  Comput.\ Phys.\ Commun.\  {\bf 134} (2001) 335,
  arXiv:hep-ph/0009029.

\bibitem{Gehrmann:2005pd}
  T.~Gehrmann, T.~Huber and D.~Maitre,
  Phys.\ Lett.\  B {\bf 622} (2005) 295,
  arXiv:hep-ph/0507061.

\bibitem{Smirnov:2008iw}
  A.~V.~Smirnov,
  JHEP {\bf 0810} (2008) 107,
  arXiv:0807.3243 [hep-ph].
%


\end{thebibliography}
\end{document}